\documentclass[final]{cvpr}

\usepackage{times}
\usepackage{epsfig}
\usepackage{graphicx}
\usepackage{amsmath}
\usepackage{amssymb}
\usepackage{booktabs}
\usepackage{caption}
\usepackage{tabularx}
\usepackage[usenames]{color}
\usepackage{indentfirst}
\usepackage{pbox}
\usepackage{makecell}
\usepackage[pagebackref=true,breaklinks=true,colorlinks,bookmarks=false]{hyperref}

\DeclareMathOperator*{\argmin}{arg\,min} 
\DeclareMathOperator*{\softmax}{softmax}

\def\etal{et al.}

\newcommand{\cmmnt}[1]{}

\def\cvprPaperID{21} 
\def\confYear{CVPR 2021}

\begin{document}

\title{D-LEMA: Deep Learning Ensembles from Multiple Annotations - \\Application to Skin Lesion Segmentation}


\author{Zahra Mirikharaji \and  Kumar Abhishek \and  Saeed Izadi  \and Ghassan Hamarneh \and
School of Computing Science, Simon Fraser University, Canada\\
{\tt\small \{zmirikha, kabhishe, saeedi, hamarneh\}@sfu.ca}
}

\maketitle

\begin{abstract}
   Medical image segmentation annotations suffer from inter- and intra-observer variations even among experts due to intrinsic differences in human annotators and ambiguous boundaries. Leveraging a collection of annotators’ opinions for an image is an interesting way of estimating a gold standard. Although training deep models in a supervised setting with a single annotation per image has been extensively studied, generalizing their training to work with datasets containing multiple annotations per image remains a fairly unexplored problem. In this paper, we propose an approach to handle annotators' disagreements when training a deep model. To this end, we propose an ensemble of Bayesian fully convolutional networks (FCNs) for the segmentation task by considering two major factors in the aggregation of multiple ground truth annotations: (1) handling contradictory annotations in the training data originating from inter-annotator disagreements and (2) improving confidence calibration through the fusion of base models' predictions. We demonstrate the superior performance of our approach on the ISIC Archive and explore the generalization performance of our proposed method by cross-dataset evaluation on the PH$^2$ and DermoFit datasets.
\end{abstract}

\section{Introduction}

The semantic segmentation task in computer vision involves partitioning an image into a set of multiple non-overlapping and semantically interpretable regions~\cite{haralick1991computer}, and this entails assigning pixel-wise class labels to the entire image, making it a dense prediction task. Segmentation is a crucial task in the visual computing pipeline and is often used to improve several downstream tasks such as classification and depth estimation~\cite{zamir2018taskonomy}. Following the seminal work of Long~\etal~\cite{long2015fully}, deep learning-based semantic image segmentation models have gained prominence because of their superior performance over traditional approaches. The majority of deep learning-based semantic segmentation models, however, rely on supervised learning of dense pixel annotations for the labels in images. State of the art supervised learning algorithms rely upon training using large volumes of data to yield acceptable results, and previous work has shown the importance of sufficient annotated data for visual tasks~\cite{oquab2014learning,huh2016makes,sun2017revisiting}. Particularly, Sun~\etal~\cite{sun2017revisiting} showed that the performance of segmentation models in terms of overlap based measures exhibits a logarithmic relationship with the amount of training data used for representation learning for semantic segmentation.

\begin{figure*}[!htbp]
    \begin{center}
    \includegraphics[width=\linewidth]{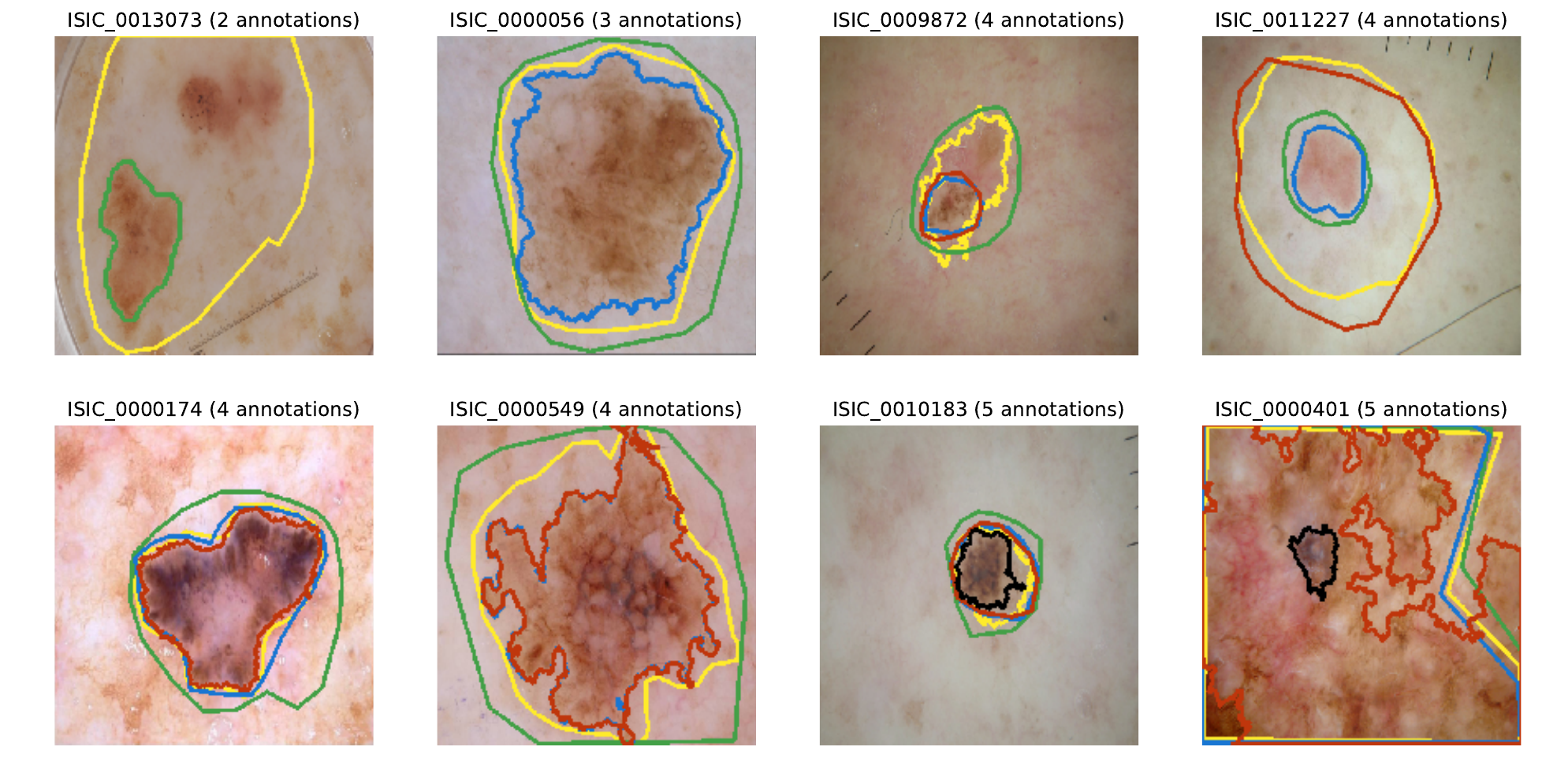}
\end{center}
    \caption{Sample skin lesion images from the ISIC Archive which contain multiple lesion boundary annotations (denoted by different colors).}
    \label{fig:ISICArchive_samples}
\end{figure*}

Collecting ground truth annotations for semantic segmentation is considerably more expensive than doing so for other visual tasks such as classification and object detection because of the dense annotations involved. While this can partly be ameliorated by crowd-sourcing the annotation process to non-experts, the presence of multiple object classes in a scene, coupled with factors such as illumination, shading, and occlusion, makes delineating the exact object boundaries an ambiguous and tedious task, leading to inter-annotator disagreements. The presence of multiple annotations (Figure~\ref{fig:ISICArchive_samples}) further leads to the challenge of deciding upon an ideal ground truth against which the model's performance is assessed. Moreover, there exists a tradeoff between the precision and the generalizability of an `ideal' segmentation ground truth, since aoverly precise delineation may not be reflective of the typical uncertainty encountered in practice when localizing the boundary~\cite{warfield2004simultaneous}. A similar trade-off exists between the quality and the efficiency of these annotations: High quality dense annotations, although useful, take up more time to collect than relatively less informative approximate annotations (e.g., bounding boxes or simplified polygons). These problems are exacerbated further for medical images since medical imaging datasets with accurate pixel-level annotations are much smaller than their natural image counterparts~\cite{taghanaki2020deep}, which can be attributed to the high cost associated with expert annotations, the difficulty in quantifying a true reference standard, the laborious nature of making dense annotations, which is even more difficult for 3D medical image volumes, and patient data privacy concerns. To add to this, the manual annotation of anatomical regions of interest can be very subjective and presents considerable inter- and intra-annotator disagreements even amongst experts across multiple medical imaging modalities~\cite{vorwerk2009delineation,fu2014interrater,taghanaki2018segmentation,ribeiro2019handling,goyal2020skin}, making it difficult to converge on a single gold standard annotation for model training and evaluation.

One of the seminal works on comparing a segmentation model's performance by comparing against a collection of (human-annotated) segmentations is that proposed by Warfield~\etal~\cite{warfield2004simultaneous}, where they proposed an expectation maximization algorithm for the simultaneous truth and performance level estimation (STAPLE). Given a collection of segmentation masks, STAPLE generates a probabilistic estimate of the true segmentation mask as well as the segmentation performance of each of the segmentations in the collection. This was followed by several other extensions of STAPLE which addressed its limitations such as susceptibilities to large variations in inter-annotator uncertainty and annotator performance~\cite{biancardi2009tesd,kamarainen2012combining,langerak2010label,li2011estimating}. 

More recently, Mirikharaji~\etal~\cite{mirikharaji2019learning} showed that leveraging different levels of annotation reliability, using spatially-adaptive reweighting while learning deep learning based segmentation model parameters, helps improve performance, and demonstrated superior segmentation accuracy using a large number of low quality, `noisy' annotations along with only a small fraction of precise annotations. Hu~\etal~\cite{hu2019supervised} used a modified probabilistic U-Net~\cite{kohl2018probabilistic} model to generate quantifiable aleatoric and epistemic uncertainty estimates for segmentation using a supervised learning framework which modeled inter-annotator variability as aleatoric uncertainty ground truth. Ribeiro~\etal~\cite{ribeiro2019handling} proposed an approach to improve inter-annotator agreement by conditioning the segmentation masks using morphological image processing operations (opening and closing), convex hulls and bounding boxes to remove details specific to any single particular annotator. They argue that the conditioning could be deemed as denoising operations, removing the annotator specific details from the segmentation masks. The same authors then proposed to train their segmentation model on a subset of the images, derived by filtering out all samples whose mean pairwise Cohen's kappa score was less than 0.5, thus using only those segmentations which largely agree between annotators~\cite{ribeiro2020less}.

Despite the obvious benefits of improving segmentation performance, it is also crucial to analyze the predictive uncertainty of deep networks in medical image segmentation. In machine learning, the uncertainty has been classified into aleatoric and epistemic types. The aleatoric, which reflects the inherent noise in the data, has been estimated using a second auxiliary output in the network~\cite{kendall2017uncertainties}. Bayesian neural networks (BNNs) have adopted Monte Carlo (MC) dropout~\cite{gal2016dropout} to reflect the epistemic uncertainty associated with the network parameters. Thanks to their simplicity, MC dropout uncertainty estimation has been studied in the context of general semantic segmentation~\cite{kendall2015bayesian} as well as medical image segmentation~\cite{kwon2020uncertainty,sedai2018joint}. However, the uncertainty estimates obtained using MC dropout tend to be miscalibrated, i.e., they do not correspond well with the model error~\cite{laves2019well}. Recently, there have been efforts to improve the uncertainty calibration using ensemble learning. Particularly, Lakshminarayanan~\etal~\cite{lakshminarayanan2017simple} demonstrated the advantage of ensemble learning, i.e., averaging a collection of models trained from different initializations, in yielding more accurate predictive uncertainty estimates for classification and regression tasks. Mehrtash~\etal~\cite{mehrtash2020confidence} studied the performance of ensemble learning for predictive uncertainty in medical image segmentation. Particular to skin lesion segmentation, Jungo~\etal~\cite{jungo2019assessing} thoroughly studied the reliability of existing uncertainty estimation methods and showed their benefits and limitations~\cite{jungo2019assessing}.

\par 

Deep neural networks have been shown to potentially overfit to noisy labels~\cite{zhang2016understanding} and our motivation for this work is to avoid single annotator bias~\cite{lampert2016empirical}. Therefore, we seek training deep segmentation models to learn from multiple annotations as available instead of discarding some annotations.
Rather than selecting a subset of images to learn from Ribeiro et al.~\cite{ribeiro2020less}, we instead propose a generalized approach of annotation weighting by leveraging different groups of consistent annotations in an ensemble method towards efficiently learning from all available annotations. We also utilize uncertainty estimates~\cite{kendall2017uncertainties,lakshminarayanan2017simple} in an ensemble learning framework to improve predictive uncertainty and calibration confidence in the final prediction.

\noindent\textbf{Contribution claims:} 
We consider two major factors in the aggregation of multiple ground truth annotations: 
(1) handling contradictory annotations in the training data originating from inter-annotator disagreements, and 
(2) improving the model’s confidence calibration through deep ensembling.
Our hypothesis is that given a new image, leveraging different experts’ skills independently and fusing them in an ensemble model, while considering their estimated uncertainty, makes for a more reliable final prediction.

\begin{figure*}[t!]
 \begin{center}
   \includegraphics[width=\linewidth]{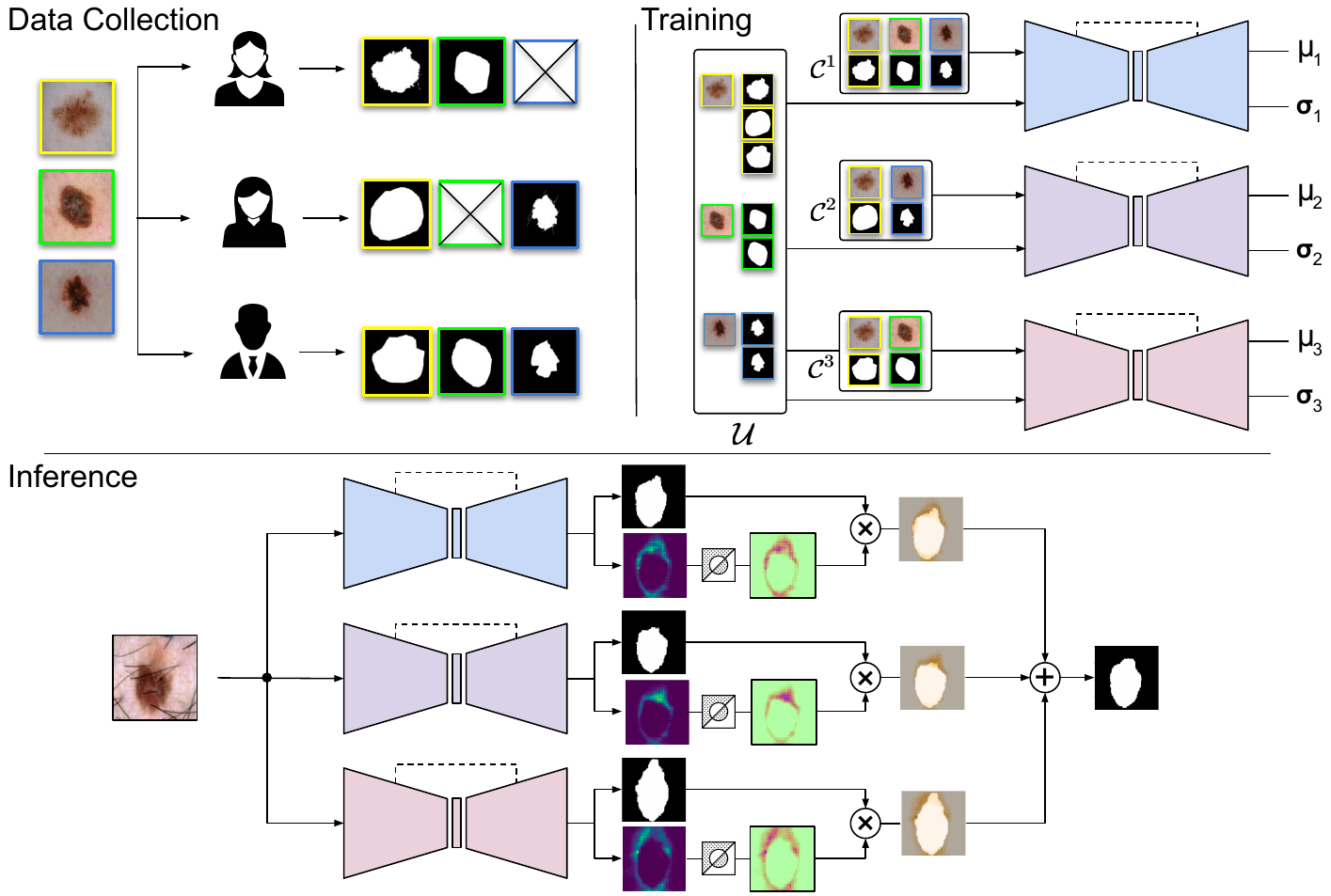}
 \end{center}
 \caption{An overview of our proposed framework for skin lesion segmentation with multiple annotations. (top left) Multiple users annotating different, potentially overlapping, subsets of the original data. (top right) Each set of non-contradictory labels is considered as ground truth and, along with the remaining annotations that are deemed potentially noisy, are used to train a different base model. (bottom) At inference, each base model's prediction, along with its estimated aleatoric uncertainty maps are fused to obtain the final prediction.}
  \label{fig:overview}
\end{figure*}

\section{Method}
\subsection{Problem Statement and Method Overview}
Let $\mathcal{X}=\{X_n\}_{1}^{N}$ and $\mathcal{Y}=\{Y_n\}_{1}^{N}$ be a set of $N$ images and segmentation ground truth masks, respectively. In a supervised learning scheme, a network is trained to learn a function $f_{\theta}: X_n \mapsto \hat{Y}_n$ parameterized by $\theta$, which maps an image $X_n$ to the corresponding estimated segmentation mask $\hat{Y}_n$. Approximating the mapping function $f_\theta$ using a single annotation per image has been well studied in the literature. However, training supervised models in the presence of multiple annotations remains largely unexplored. 

Let us assume that $K$ annotators have independently annotated different subsets of the images resulting in a set of segmentation ground truths $\mathcal{Y}=\{\{Y_{mn}\}_{m=1}^{M_n} \}_{n=1}^{N}$, where $M_n$ denotes the number of available annotations for $X_n$. Inconsistent annotations for a given image could mislead the network and substantially deteriorate the performance of the model. Let $M$ indicate the maximum number of annotations per image over the entire dataset. Instead of aggregating multiple annotations to estimate a single ground truth before the training phase, we propose to (1) learn a set of $M$ mapping functions $\mathcal{F} = \{f_{\theta^{i}}\}$ through ensembling $M$ base deep models trained over the union of available annotations and (2) minimize the confusion induced from observing multiple annotations through a spatial re-weighting scheme during training. (3) Lastly, we demonstrate that our proposed ensemble learning framework not only improves the segmentation performance but also provides a well-calibrated predictive uncertainty. Figure~\ref{fig:overview} illustrates the overview of our ensemble learning framework for skin lesion segmentation with multiple annotations.

\subsection{Detailed Method}
\noindent\textbf{Non-contradictory Subsets Selection}: To handle contradictory annotations arising from having multiple annotations per image during the training, we partition the entire dataset into $M$ disjoint subsets, denoted by \{$\mathcal{C}^i\}_{i=1}^{M}$, such that each $\mathcal{C}^i$ includes at most one unique annotation for every image. In particular, for each image, with $M_n \leq M$ annotations, we randomly assign the $M_n$ annotations to \{$\mathcal{C}^i\}_{i=1}^{M_n}$ subsets.

A naïve approach is to utilize these disjoint subsets to train individual base models independently. Even though this solution prevents exposing each ensemble base model to multiple annotations per image and encourages a diverse set of model performance, however, each disjoint set includes a small number of training samples which can adversely affect the generalization capability of individual base models. To address this issue, we combine all images along with all available annotations into a \textit{union} dataset, denoted as $\mathcal{U}$, and use it to train $M$ base networks.  Following Mirikharaji et al.~\cite{mirikharaji2019learning}, we utilize these non-contradictory subsets to assess the quality of annotations in $\mathcal{U}$. Specifically, spatially-adaptive weight maps associated with varying annotations in $\mathcal{U}$ are learned to adjust the contribution of each annotated pixel in the optimization of deep network based on its consistency with clean annotations in \{$\mathcal{C}^i\}$.

\noindent\textbf{Learning Models}: In more details, for each base model $i$, $i \in {1,...,M}$, we define a cross-entropy loss, denoted as  $\mathcal{L} = \{L_{ce}^{\mathcal{C}^i}\}$ over each non-contradictory set $\mathcal{C}^i$. We then, in a meta-learning paradigm, learn a set of spatial weight maps $
\mathcal{W}^i = \{\{W_{mn}^i\}_{m=1}^{M_n} \}_{n=1}^{N}$ for all annotations $\mathcal{U}$ based on the gradients of the cross-entropy losses with respect to the weights maps, i.e. $\nabla_{W^{i}}\mathcal{L}_{ce}^{\mathcal{C}^i}$. This way, $\mathcal{W}^i$ is optimized to cancel out the contributions of annotations inconsistent with $\mathcal{C}^i$ while optimizing the parameters for $i^{\mathrm{th}}$ base network, i.e. $\theta^i$. Mathematically: 

\begin{eqnarray}
\mathcal{W}^{i\ast}= \argmin_{\mathcal{W}^i,~\mathcal{W}^i\geqslant0}\sum_{n \in \mathcal{C}^i} L_{ce}^n(\hat{Y}_n^i,Y_{n};\theta^i(\mathcal{W}^i)).
\label{optimal_weights}
\end{eqnarray}

\noindent Note that every image in $\mathcal{C}^i$ has only one ground truth. $\mathcal{W}^i$ are encoded in $\mathcal{L}$ and they are optimized along with the network parameters $\theta^i$ for each individual base model. By integrating the information in the optimized $\mathcal{W}^i$, we can determine the degree by which a pixel-level annotation from any of annotators is considered noisy for model $i$, depending on how similar this annotation is to the annotations in $\mathcal{C}^i$. Therefore:

\begin{eqnarray}
\mathcal{L}(\hat{Y}_n^i,Y_{mn};\theta^i, W_{mn}^i)= -\sum_{q\in X_n}  W_{mnq}^i Y_{mnq}\log \hat{Y}_{nq}^i,\\ 
\hat{Y}_{nq}^i = \softmax(U_{nq}^i).~~~~~~~~~~~~~~~~~~~
\label{eqn:L}
\end{eqnarray}

\noindent\textbf{Fusion of Predictions}: Once the individual base models are trained, the final prediction of the entire ensemble for the $X_n$ is obtained by using a weighted fusion~\cite{sagi2018ensemble}, that is:

\begin{eqnarray}
    \hat{Y}_n = \sum_{i=1}^{M} \alpha^i_n\hat{Y}^i_n,
\end{eqnarray}

\noindent where $\alpha^i_n$ is the combination coefficient for prediction by model $i$. The simplest way to determine $\alpha_n^{i}$ is to consider equally weighted averaging and set them to ${1}/{M}$. Another popular technique is to set $\alpha_n^i$ coefficients according to the confidence of the model~\cite{schapire1999improved}. In this work, we explore both aggregation techniques in our experimental evaluations.

\noindent\textbf{Uncertainty-driven Aggregation}: For the uncertainty-driven aggregation of base models, we leverage aleatoric uncertainty, which models irreducible observation noise, to estimate how confident a base model is about its prediction, and utilize the confidence when combining the base models' prediction maps. Following Kendall~\etal~\cite{kendall2017uncertainties}, we approximate the aleatoric uncertainty for each pixel $q \in X_n$ by placing a Gaussian distribution over the logit space before applying a sigmoid function in the last layer and reformulate the network output as:

\begin{eqnarray}
U_{nq}^i \sim \mathcal{N}\left(f_{nq}^i, (\sigma^i_{nq})^2\right),
\end{eqnarray}
where $f_i$ and $\sigma_i$ are the network $i$ outputs.

We use the aleatoric uncertainty in two forms: (1) considering the pixel-wise uncertainty values as spatially-adaptive coefficients and (2) averaging the pixel-wise uncertainty into a single scalar image-level coefficient.


\section{Experiments}
\subsection{Data}

For training, we used the International Skin Imaging Collaboration (ISIC) Archive data~\cite{ISICArchive, codella2018skin, codella2019skin}, the largest dermoscopic public dataset with over 13,000 images, captured by diverse devices in international clinical centers. All images are 8-bit RGB color dermoscopy images. Similar to Ribeiro et al.~\cite{ribeiro2020less}, we utilized 2,223 images with more than one segmentation ground truth mask (2,094 with two, 100 with three and 36 with four and 3 with five) to train our models. We split all 2,223 images to 80\% for training and 20\% for validation. For model selection, we randomly selected which annotation to use in validation set. To create our non-contradictory annotation sets, all training data are randomly and uniformly partitioned into five groups of overlapping images but unique ground truth annotations. ISIC ground truth masks were generated using three different pipelines with different levels of border irregularities all involving a dermatologist with expertise in dermoscopy: (1) an automatic algorithm followed by an expert review; (2) a semi-automatic algorithm controlled by an expert; and (3) manually drawing a polygon by an expert. A large variation of disagreement based on Cohen’s kappa scores with the mean 0.67 is reported in Ribeiro et al.~\cite{ribeiro2019handling}. Figure~\ref{fig:ISICArchive_samples} shows some examples of skin lesion images with multiple lesion boundary annotations from this dataset.

\par

To thoroughly assess the segmentation performance of our proposed ensemble framework, we leveraged three publicly available datasets in our evaluations. All the images in the used datasets are resized into $96 \times 96$ pixels and normalized using the per-channel mean and standard deviation across the entire dataset. A brief description of these test datasets are provided as follows: \par
\begin{itemize}
\item[$\bullet$] \textbf{ISIC}: Ribeiro et al.~\cite{ribeiro2020less} randomly selected a test set of 2,000 images with just one segmentation ground truth from ISIC Archive. We used the exact set in our experimental evaluations for fair comparisons.
\item[$\bullet$] \textbf{PH$^2$}: The PH$^2$ (Pedro Hispano Hospital) dataset contains 200 8-bit RGB color dermoscopic images~\cite{mendoncca2013ph}. All images are acquired under the same condition using Tuebinger Mole Analyzer system at 20$\times$ magnification.
\item[$\bullet$] \textbf{DermoFit}: This dataset has 1300 8-bit RGB color clinical images~\cite{ballerini2013color}. The images are captured with a Canon EOS 350D SLR camera at the same distance from the lesion under controlled lighting conditions.
\end{itemize}

\subsection{Base Models and Implementation Details}
Our architecture is an encoder-decoder architecture with residual and skip connections transferring the information in the encoder modules to the corresponding decoder modules~\cite{chaurasia2017linknet}. Since the images in our training dataset are paired with at most five annotations ($M=5$), our ensemble framework consists of five base deep neural networks. Each network outputs two spatial maps in the last layer: the dense segmentation prediction and the predicted aleatoric uncertainty map. In training the aleatoric loss, 10 Monte Carlo samples from logits are taken. 
Stochastic gradient descent with an initial learning rate of $10^{-4}$ is used to optimize the network parameters. The batch size for optimizing the spatial weight maps and network parameters is 64 and 2. The momentum and weight decay are set to 0.99 and 5 $\times 10^5$, respectively.

\subsection{Results}
\begin{table*}[!t]
\renewcommand{\arraystretch}{1.2}
\setlength{\tabcolsep}{6pt}
\centering
\caption{Comparing the segmentation performance based on Jaccard index reported in percent ($\%$ $\pm$ standard error) on three datasets.}
\begin{center}
\begin{tabular}{|c|c|c|c|c|}
\hline
\textbf  & Method & \text~ISIC Archive~\cite{ISICArchive}~& ~PH$^2$~\cite{mendoncca2013ph}~& ~DermoFit~\cite{ballerini2013color}~\\
\hline
\hline
~A~ & baseline & 68.00 $\pm$ 0.56 & 81.30 $\pm$ 0.77 & 70.30 $\pm$ 0.54\\
\hline
~B~ & model 0& 69.22 $\pm$0.53 & 82.82 $\pm$ 0.75 & 72.57 $\pm$ 0.50 \\
\hline
~C~ & model 1& 69.75 $\pm$ 0.55 & 82.40 $\pm$ 0.75 & 71.05 $\pm$ 0.55\\
\hline
~D~ & model 2 & 70.33 $\pm$ 0.52 & 83.46 $\pm$ 0.74& 72.80 $\pm$ 0.51 \\
\hline
~E~ & model 3 & 70.37 $\pm$ 0.51 &  83.31 $\pm$ 0.70 & 73.04 $\pm$ 0.53\\
\hline
~F~ & model 4 & 69.73 $\pm$ 0.52 & 82.29 $\pm$ 0.72 & 70.87 $\pm$ 0.48\\
\hline
~G~ & \makecell{equally weighted fusion (ours)} & \textbf{72.11$\pm$ 0.51} & 84.96$\pm$ 0.73  & 74.22$\pm$ 0.51\\
\hline
~H~ & \makecell{pixel-level confidence (ours)} & 71.46$\pm$ 0.49 &  84.52$\pm$ 0.74 & 73.91$\pm$ 0.53\\
\hline
~I~ & \makecell{image-level confidence (ours)} & \textbf{72.08$\pm$ 0.49} &  \textbf{85.20 $\pm$ 0.70} & \textbf{74.33$\pm$ 0.50}\\
\hline
~J~ & less is more~\cite{ribeiro2020less} & 69.20 & 81.25 & 72.55 \\
\hline
\end{tabular}
\end{center}
\label{seg_performance}
\end{table*}
\begin{table*}[!h]
\renewcommand{\arraystretch}{1.2}
\setlength{\tabcolsep}{8pt}  
\centering
\caption{Comparing predictive uncertainty based on negative log-likelihood (NLL) and Brier score (Br) on three datasets. Lower NLL and Br values correspond to a better predictive uncertainty estimate.}
\begin{center}
\begin{tabular}{|c|c|c|c|c|c|c|c|}
\hline
\multicolumn{2}{|c|}{Dataset}& \multicolumn{2}{|c|}{ISIC Archive} & \multicolumn{2}{|c|}{PH$^2$} & \multicolumn{2}{|c|}{DermoFit}\\
\hline
\hline
\multicolumn{2}{|c|}{Method} & ~NLL~& ~Br~ & ~NLL~& ~Br~ & ~NLL~& ~Br~\\
\hline
~A~ & \makecell{MC dropout model 0} & 0.073 & 0.019 & 0.166 & 0.048 & 0.272 & 0.082 \\
\hline
~B~ & \makecell{MC dropout model 1}& 0.075 & 0.020 & 0.151 & 0.044 & 0.310 & 0.099 \\
\hline
~C~ & \makecell{MC dropout model 2}& 0.075 & 0.019 & 0.149 & 0.044 & 0.283 & 0.087 \\
\hline
~D~ & \makecell{MC dropout model 3}& 0.078 & 0.020 & 0.152 & 0.042 & 0.291 & 0.091 \\
\hline
~E~ & \makecell{MC dropout model 4}& 0.075 & 0.019 & 0.155 & 0.045 & 0.312 & 0.100 \\
\hline
~F~ &  \makecell{deep ensemble (ours)} & \textbf{0.070} & \textbf{0.018} & \textbf{0.144} & \textbf{0.041} & \textbf{0.254} & \textbf{0.078} \\
\hline
\end{tabular}
\end{center}
\label{pred_uncertainty}
\end{table*}
Table~\ref{seg_performance} compares the segmentation performance of our baseline models as well as the individual base models, across different prediction fusion schemes, using the Jaccard index. To train the baseline model, for every image in the training batch, we randomly select which ground truth to use when optimizing the loss function (row A). While it is interesting to consider each annotator separately and evaluate their performance, the assignments between annotators and ground truth are not stated in the ISIC Archive dataset. Instead, we evaluate the performance of each base model trained on non-contradictory annotations simulating an expert knowledge (rows B to F). In addition, we compare the performance of our proposed method against the work of Ribeiro~\etal~\cite{ribeiro2020less} where a subset of samples with small annotator disagreements is taken into account during the training.

For the fusion stage, we examine three approaches as listed below:
\begin{itemize}
    \item[$\bullet$] \textbf{Uniformly weighted fusion}: The predictions from the base models are combined by averaging the output probabilities.
    \item[$\bullet$] \textbf{Pixel level confidence-based fusion}: The predictions from the models are fused using normalized confidence spatial maps computed by inverting the predicted aleatoric outputs. 
    \item[$\bullet$] \textbf{Image level confidence-based fusion}: The aleatoric uncertainty maps are aggregated into an image level aleatoric scalars and the predictions of the base models are combined based on the image-level normalized confidence scalars computed by inverting the uncertainty scalars.
\end{itemize}

Our results demonstrate that leveraging all available annotations effectively in an ensemble framework consistently improves the performance of the segmentation performance both in a held-out test set and over two other distinct datasets. Looking into different variants of our deep ensemble method, it is evident that aggregating the aleatoric uncertainty into the image-level scalar and leveraging them in the fusion stage (row H) either outperforms or exhibits competitive performance against the uniform averaging scheme (row G).
\begin{figure*}[h]
 \begin{center}
   \includegraphics[width=0.8\linewidth]{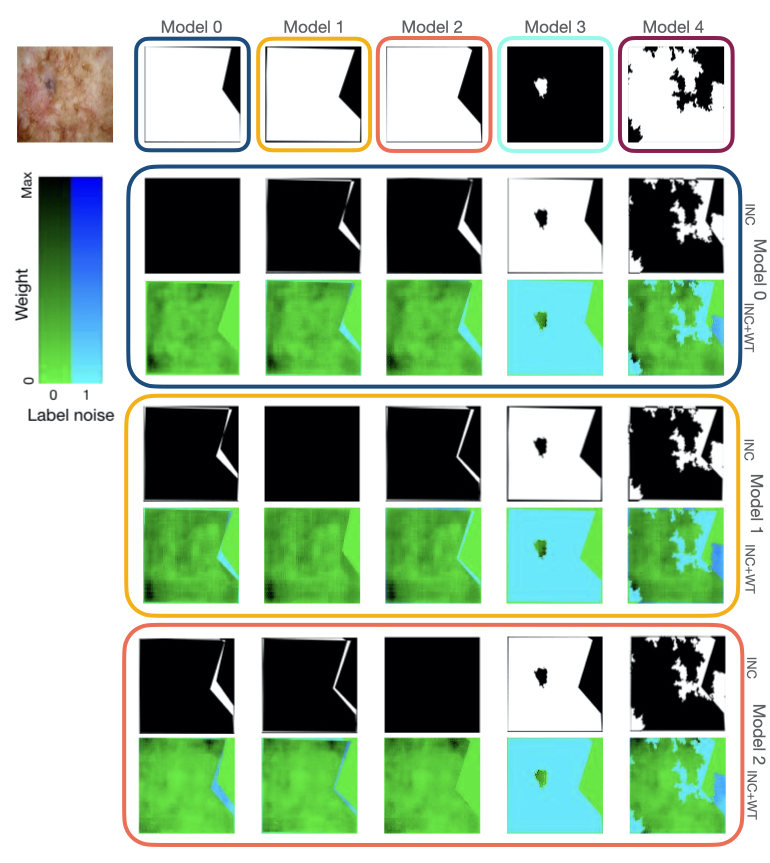}
 \end{center}
 \vspace{-1em}
 \caption{Qualitative evaluation of weighting matrices: (first row) a sample training image and trusted annotations in base models 0 to 4. (second row) inconsistency maps (INC) between the trusted ground truth in Model 0 and other ground truth annotations. (third row) learned weight maps in iteration 100K overlaid over the inconsistency maps (INC+WT). Color-coded boxes indicates the change when the trusted annotations in base models 0, 1 and 2 are different.}
  \label{fig:w_maps1}
\end{figure*}

\begin{figure*}[h]
 \begin{center}
   \includegraphics[width=0.8\linewidth]{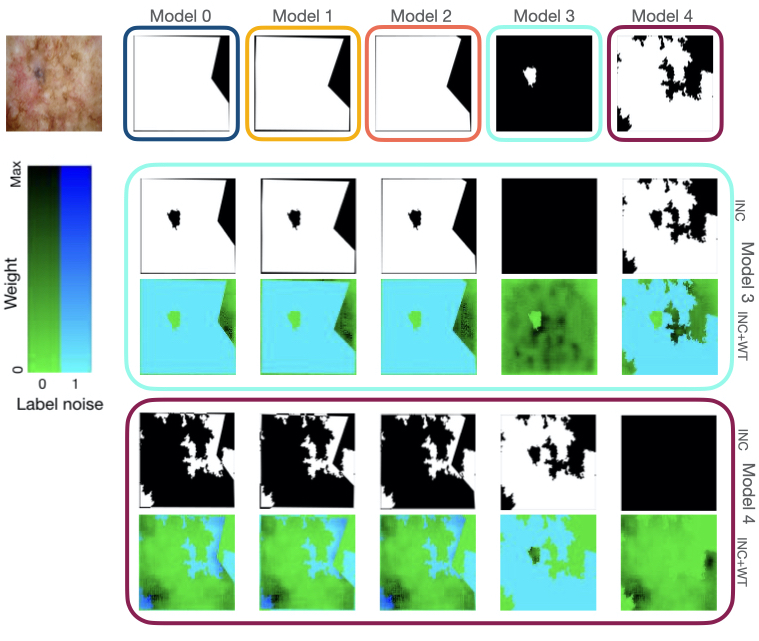}
 \end{center}
 \vspace{-1em}
 \caption{Qualitative evaluation of weighting matrices: (first row) a sample training image and trusted annotations in base models 0 to 4. (second row) inconsistency maps (INC) between the trusted ground truth in Model 3 and other ground truth annotations. (third row) learned weight maps in iteration 100K overlaid over the inconsistency maps (INC+WT). Color-coded boxes indicate the changes when the trusted annotations in base models 3 and 4 are different.}
  \label{fig:w_maps2}
\end{figure*}

While modeling predictive uncertainty in clinical applications without a ‘real’ gold standard is helpful in decision making, miscalibrated uncertainty with overconfident predictions leads to an unreliable outcome. To evaluate the calibration quality of our ensemble annotation aggregation against Bayesian FCNs, we implemented Bayesian epistemic uncertainty using dropout for each base model.
Similar to Bayesian SegNet~\cite{kendall2015bayesian}, we added five dropout layers in the central part of the encoder and the decoder after each convolutional layer. Dropout probability is set to 0.3 and they are kept active at the inference time. Fifteen feed-forwards are executed to perform MC sampling and the output mean is considered as the final segmentation prediction.

To evaluate the quality of the predictive uncertainty, we use two widely used metric in the literature~\cite{lakshminarayanan2017simple,gal2016dropout}; negative log-likelihood (NLL) and Brier score (Br). Given a segmentation network with sigmoid non-linearity in the output layer, NLL and Br for $X_n$ are calculated as follows: \\
\begin{equation}
    NLL = \frac{-1}{|X_n|} \sum_{q\in X_n} Y_{nq}\log\hat{Y}_{nq} + (1-Y_{nq})\log (1-\hat{Y}_{nq}) \\   
\end{equation}
\begin{equation}
    Br =  \frac{1}{|X_n|} \sum_{q\in X_n}  [Y_{nq}- \hat{Y}_{nq}]^2    
\end{equation}

Consistent with prior studies on deep ensembling~\cite{lakshminarayanan2017simple,mehrtash2020confidence}, Table~\ref{pred_uncertainty} indicates that our annotation aggregation ensemble with five base models consistently improves the confidence calibration and predictive uncertainty for three datasets in comparison to modeling epistemic uncertainty by MC dropout.

The spatially adaptive weight maps for model $i$, $\mathcal{W}^i$, are learned to prevent penalizing the pixels whose feature maps are similar to the feature maps of data in $\mathcal{C}^i$ while their gradient direction is not similar to the direction of loss gradient on annotations in $\mathcal{C}^i$. To qualitatively evaluate matrices $\mathcal{W}^i$, in Figures~\ref{fig:w_maps1} and~\ref{fig:w_maps2}, we overlay the learned weight maps, in training iteration 100K, over the inconsistency maps (absolute differences of ground truth masks). 
Looking into the color-coded boxes shows how the location of the cyan pixels matches the inconsistency maps (zero or very close to zero weights are assigned to inconsistent annotated pixels), which results in exclusively leveraging the experts knowledge in $\mathcal{C}^i$ when learning $\mathcal{\theta}^i$.

\section{Conclusion}
Approaches to train deep segmentation models do not trivially generalize to datasets with multiple image annotations. We propose an ensemble paradigm to deal with discrepancies in segmentation annotations. A robust-to-annotation-noise learning scheme is utilized to efficiently leverage the multiple experts’ opinions toward learning from all available annotations and improve the generalization performance of deep segmentation models. The quality of predictive uncertainty in clinical applications without true gold standards is critical. Our model captures two types of uncertainty, aleatoric uncertainty modeled in the training loss function and epistemic uncertainty modeled in the ensemble framework to improve confidence calibration. 

\noindent\textbf{Acknowledgments.} We gratefully thank the Natural Sciences and Engineering Research Council (NSERC) of Canada for funding and the NVIDIA Corporation for the donation of the Titan X GPU used for this research.

{\small
\bibliographystyle{ieee_fullname}
\bibliography{references}
}

\end{document}